\documentclass[aps,prd,showpacs,nofootinbib]{revtex4}
\usepackage{dcolumn}
\usepackage{amsmath,amssymb,setspace,graphicx,subfigure,epsfig}
\usepackage{amsfonts}
\usepackage{latexsym}
\usepackage{epsfig}
\usepackage{color}
\usepackage{nicefrac}

\def\fNL{f_{NL}}

\def\be{\begin{equation}}
\def\ee{\end{equation}}
\def\bea{\begin{eqnarray}}
\def\eea{\end{eqnarray}}
\def\MPl{M_{\rm Pl}}

\def\P{{\mathcal{P}}}

\def\esigsig{\eta_{\sigma\sigma}}

\def\e{\epsilon}

\def\({\left(}
\def\){\right)}
\def\nn{\nonumber}

\def\0{^{(0)}}
\def\1{^{(1)}}
\def\2{^{(2)}}

\def\MPl{M_{Pl}}
\def\P{{\cal P}}

\begin{document}
\title{Constraints on generating the primordial curvature perturbation and non-Gaussianity from
instant preheating}
\author{Christian T. Byrnes}
\email{C.Byrnes@thphys.uni-heidelberg.de}
\affiliation{Institut f\"ur Theoretische Physik, Universit\"at Heidelberg, Philosophenweg 16, 69120
Heidelberg, Germany}

\pacs{98.80.Cq }

\hfill HD-THEP-08-24

\begin{abstract}
We analyse models of inflation in which isocurvature perturbations present during inflation are
converted into the primordial curvature perturbation during instant preheating. This can be due to
an asymmetry
between the fields present either during inflation or during preheating. We consider all the
constraints that the model must satisfy in order to be
theoretically valid and to satisfy observations. We show that the constraints are very tight in all
of the models proposed and special initial conditions are required for the models to work. In the
case where the symmetry is strongly broken during inflation the non-Gaussianity parameter $\fNL$ is
generally large and negative.
\end{abstract}

\maketitle


During multiple field inflation there are two types of perturbations present. The perturbations
parallel to the background trajectory are adiabatic perturbations, while those perpendicular are
isocurvature perturbations. Various methods of converting the inflationary isocurvature
perturbations into the primordial curvature perturbation have been considered. Here we consider
instant preheating, a very fast and efficient method of converting the inflaton fields into
radiation at the end of inflation \cite{felder}. If there is an asymmetry between the fields either
during slow-roll inflation, or during instant preheating then this can convert the initial
isocurvature perturbation into the primordial curvature perturbation. Several specific models for
doing this have been proposed in the literature
\cite{Kolb:2004jm,byrnes,Matsuda:2006ee,Battefeld:2007st,Riotto:2008gs}. However the parameter
constraints have in
most cases not been calculated and we show that it is not easy to realise this scenario.
Specifically we require three key conditions to be simultaneously satisfied, that the primordial
curvature perturbation is of the required amplitude to match observations, that is is generated from
the inflationary isocurvature perturbation and that instant preheating is efficient so that this
scenario is valid. We show that satisfying all of these conditions requires special initial
conditions tuned so that the background trajectory is nearly parallel to one of the scalar field
axes throughout inflation. For each of the models we also calculate the other observable quantities
such as the spectral index and amount of non-Gaussianity in the allowed region of parameter space
which we show can be large.


We use the simple quadratic potential of multiple field chaotic inflation, but include the
effects of the isocurvature perturbations during inflation. 
After introducing some notation and formula in the next section we
introduce instant preheating in Sec.~\ref{ip:sec:ip} and discuss how this can be used
to convert the inflationary isocurvature perturbations into the primordial curvature
perturbation. We then discuss three specific models, nearly symmetric inflation in
Sec.~\ref{ip:sec:ko}, highly non-symmetric inflation in Sec.~\ref{ip:sec:ma} and symmetric
inflation followed by non-symmetric preheating in Sec.~\ref{ip:sec:by}. Finally we draw our
conclusions in Sec.~\ref{ip:sec:conclusion} with a general discussion of the reason why it is hard
to make these models work and the reasons for the tuning of the initial conditions that we
require. We also summarise the future observations that can distinguish between or rule out these
models.

\section{Background theory}

To give rise to the observed anisotropies in the microwave background sky and
large-scale structure in our Universe today, the scalar field
perturbations during inflation must produce density perturbations
in the radiation dominated era after inflation. These primordial
perturbations are usefully characterised in terms of the
dimensionless density perturbation on spatially-flat
hypersurfaces. For linear perturbations we define
\begin{equation}\label{zetadefinition}
\zeta = - \frac{H\delta\rho}{\dot\rho} \,.
\end{equation}
This is equivalent to the perturbed expansion, $\delta N=H\delta
t$, up to a uniform-density hypersurface some time after
inflation.

Because we are interested in perturbations on scales very much larger than the Hubble
scale, the local expansion, $N$, is given in terms of the background solution for the
local values of the scalar fields on an initial spatially-flat hypersurface during
inflation. This is known as the ``separate universes'' approach \cite{WMLL}. Thus for one
or more scalar fields, $\phi_I$, during inflation, we have \cite{starob85,Sasaki:1998ug}
\begin{equation}
\zeta = \sum_I \frac{\partial N}{\partial\phi_I} \delta\phi_I \,.
\end{equation}
We treat the fields as uncorrelated at Hubble-exit, which is valid to at least first-order in
slow roll \cite{vanTent:2003mn,byrnes2}, and use $\P_\phi=\P_{\phi_1}=\P_{\phi_2}=(H_*/(2\pi))^2$
. The power spectrum of primordial density perturbations $\zeta$ can be
written as a sum of contributions from the power spectra of individual fields,
\begin{equation}
{\cal P}_\zeta = \sum_{I=1}^2 \left( \frac{\partial N}{\partial\phi_I} \right)^2 {\cal
P}_\phi\,.
\end{equation}
This can then be split into a contribution from inflaton
perturbations, $\delta\sigma$, and the orthogonal isocurvature
perturbation during inflation $\delta \chi$
\begin{equation}
\label{Pzetasplit}
{\cal P}_\zeta = \left( \frac{\partial N}{\partial\sigma}
\right)^2 {\cal P}_{\phi} + \left(
\frac{\partial N}{\partial\chi}\right)^2 {\cal P}_{\phi} \,.
\end{equation}

Lyth and Rodriguez \cite{LR} have pointed out that the
extension of this result to second-order also allows one to
calculate the non-Gaussianity of primordial perturbations due to the
non-linear dependence of the expansion after Hubble exit on the
initial field values. In the case that a single field direction $\chi$ generates the primordial
curvature perturbation the local non-linearity parameter defined by
$\zeta=\zeta_G+3/5\fNL\zeta_G^2$ is given by
\bea\label{fNLformula} \fNL=\frac{5}{6}\frac{N_{\chi\chi}}{N_{\chi}^2}, \eea 
where $\zeta_G$ is the linear, Gaussian part of the curvature perturbation and
$N_{\chi}=\partial N/(\partial \chi_*)$ etc.

Non-Gaussianity from adiabatic field fluctuations in single-field inflation are small
(first-order in slow-roll parameters \cite{Maldacena}) and remain small at Hubble exit
for multiple field inflation \cite{Seery:2005gb}. In most cases the curvature perturbation remains
 nearly Gaussian during slow-roll inflation
\cite{Rigopoulos:2005us,Vernizzi:2006ve,Battefeld:2006sz,Choi:2007su,Yokoyama:2007uu} but for a
special trajectory in some models it can become large \cite{byrnes5}. In this paper we only consider
fields with a canonical kinetic term. In models of inflation with non-canonical kinetic terms (such
as k-inflation or Dirac-Born-Infeld inflation) the scalar field fluctuations at Hubble exit may be
strongly non-Gaussian, see for example 
\cite{Chen:2006nt,Gao:2008dt,Langlois:2008qf,Arroja:2008yy}.

\subsection{Generating the primordial curvature perturbation}

We will consider three models for generating the curvature perturbation from preheating.
We can classify them according to how they break the symmetry of the inflaton fields
either before or during preheating. All  the models we will consider, when specialising
to the case of two fields, have the potential and interaction lagrangian given by
\bea\label{ip:V2field}
V(\phi_1,\phi_2)&=&\frac{1}{2}\left(m_1^2\phi_1^2+m_2^2\phi_2^2\right)\,, \\
\label{ip:g2field} \mathcal{L}_{\rm{int}}&=&\frac12\left(g_1^2\phi_1^2+g_2^2\phi_2^2\right)\psi^2\,.
\eea

If there is complete symmetry between the field both during and after inflation (i.e.~if $m=m_1=m_2$
and $g=g_1=g_2$) then only the adiabatic perturbations affect the expansion history of the universe
at first order. The background trajectory will be a straight line going through the minimum of the
potential. In this case the power spectrum of the fluctuations will be given by \cite{byrnes}
\begin{equation}
\label{ip:Pzetainf} {\cal P}_{\zeta,{\rm inf}} \simeq
\frac{4}{3\pi} \left( \frac{m}{M_{Pl}}
\right)^2 N^2,
\end{equation}
where $N\simeq60$ is the number of $e$-foldings from Hubble exit during inflation till the end of
inflation. Given the COBE normalisation, $\mathcal{P}_\zeta\simeq2\times10^{-9}$
\cite{Bunn:1996py,wmap5}, then for $m\simeq 10^{-6} M_{\rm Pl}$ fluctuations in the
inflaton field produce primordial density perturbations of the observed magnitude. On the
other hand for $m<10^{-6} M_{\rm Pl}$ the inflaton perturbations are below the
observational bound and we would require an additional contribution from isocurvature
perturbations to generate the primordial density perturbations. The primordial density
perturbations due to inflaton fluctuations will have the spectral tilt
$n_\zeta-1\simeq -0.03$.


If the symmetry between the two inflaton fields is broken either during inflation, by having
unequal mass terms, or after inflation by having unequal couplings to the preheat field then the
isocurvature perturbations, which are inevitably present during inflation with more than one light
scalar field, may be converted into a curvature perturbation. We will consider in depth three
different models for breaking the symmetry between the two fields.

The first two models we consider break the symmetry during inflation, $m_1\neq m_2$
but have symmetric preheating $g_1=g_2$. In one model the potential is nearly symmetric,
$m_1\simeq m_2$ \cite{Kolb:2004jm} while in the second model there is a strong mass hierarchy
$m_1\gg m_2$ \cite{Matsuda:2006ee}. In the final model we will consider the potential is
symmetric, $m_1=m_2$ but there is a strong symmetry breaking after inflation, $g_1\ll
g_2$ \cite{byrnes}.

For each model we will find the parameter values required so that the model satisfies
observations. Specifically we require
\begin{itemize}
\item The inflationary isocurvature perturbations generate the primordial curvature
perturbation.
\item Instant preheating is effective, $\rho_{\psi}/\rho_{\sigma}=\mathcal{O}(1)$.
\item The power spectrum satisfies the COBE normalisation, $\P_{\zeta}=2\times 10^{-9}$.
\item The non-Gaussianity is not too large, $|f_{NL}|<100$.
\end{itemize}

Assuming that the inflationary adiabatic perturbations are negligible the curvature
perturbation generated during instant preheating  from the inflationary isocurvature
perturbations (in the spatially flat gauge) is given by \cite{Kolb:2004jm}
\bea\label{ip:zeta} \zeta=\alpha\frac{\delta n_{\psi}}{n_{\psi}}\,, \eea
where $\alpha$ is a dimensionless quantity of order unity. If the $\omega$ particles are
massive non-relativistic particles then $\alpha=1/3$, while if the $\omega$ particles are
light and act like radiation then $\alpha=1/4$.

\section{Instant preheating}\label{ip:sec:ip}

We will consider a model of instant preheating at the end of inflation where the inflaton
field's energy density can be transferred to a preheat field during the first oscillation
of the inflaton field \cite{felder}. This simplifies the calculation of particle
production as a function of the initial field values, making it possible to estimate
analytically the effect on preheating of the isocurvature perturbations during inflation.

In this scenario the preheat particles $\psi$ are created through an interaction term with the inflaton field $\sigma$
\bea\label{ip:interaction} \mathcal{L}_{\rm{int}}=-\frac12g^2\sigma^2\psi^2\,,  \eea
during a brief period around $\sigma=0$, when the effective mass of the $\psi$ particles
are at their minimum. Here we are assuming the effective mass of the $\psi$ particles is
negligible. We use the notation $\sigma$ for the inflaton field in order to be general (see Sec.~\ref{ip:sec:by}), if there are several fields of equal mass then the adiabatic field $\sigma$ effectively acts as a single inflaton field. The energy density in the $\psi$-field is then ``fattened" by the coupling to
the inflaton field as $\sigma$ rolls back up the potential because of their effective
mass, $m_\psi^2=g^2\sigma^2$. Through a Yukawa interaction
$\lambda\psi\omega\overline{\omega}$ the preheat field can then decay into $\omega$
particles, decaying most rapidly when the $\psi$ particles reach their maximum effective
mass, as the inflaton field reaches its maximum. Depending on the coupling constants this
process may be so efficient that all further decay of the inflaton field can be neglected
\cite{reheating,felder}.

Particle creation in the preheat field first occurs when the adiabatic condition fails,
$|\dot{m}_\psi|=m_\psi^2$, shortly before the inflaton first passes through its minimum.
Denoting this time with a ``$0$'', so that
e.g.~$\sigma_0=\sigma(t_0)$, the adiabatic condition is broken when
\begin{equation}
|\dot{\sigma}_0|=|\sigma_0|^{2}g\,.
\end{equation}
Numerically we find $|\dot\sigma_0|\simeq m\Phi$, where $\Phi\simeq 0.07\MPl$
would be the amplitude of the first oscillation of the inflaton field after slow roll
ends \cite{felder}, if there was no transfer of energy to the preheat field. This result
follows from conservation of energy if we neglect the Hubble expansion, which is
justified because the oscillation of the field occurs on a shorter timescale than
the Hubble scale.

The time interval during which particle creation takes place is
\begin{equation}
\Delta t_0\sim\frac{\sigma_0}{|\dot{\sigma}_0|}
=|\dot{\sigma}_0|^{-1/2}{g}^{-1/2}\,.
\end{equation}
After the inflaton field has passed through the origin for the first time the occupation
number of the $\psi$ field with wavenumber $k$ is \cite{felder}
\begin{eqnarray}
n_k=\exp\left(-\pi\Delta t_0^2 k^2\right)\,.
\end{eqnarray}
This formula can be extended to the case where the $\psi$ particles have a bare mass, the
more general result is then \cite{reheating,felder}
\begin{eqnarray}
n_k=\exp\left(-\pi\Delta t_0^2\left(k^2+m_{\psi,\rm{bare}}^2\right)\right)\,.
\end{eqnarray}
Continuing to include the bare mass, $m_{\psi,\rm{bare}}$, we integrate $n_k$ to give the
total number density of $\psi$-particles produced
\begin{eqnarray}\label{ip:npsi}
n_\psi &=&\frac1{(2\pi)^3}\int^\infty_0d^3k\,n_k =(2\pi\Delta
t_0)^{-3} \exp\left(-\pi\Delta t_0^2m_{\psi,\rm{bare}}^2\right) \nn \\
&=&\frac{\left(g|\dot{\sigma}_0|\right)^{3/2}}{(2\pi)^3} \exp\left(-\pi
\frac{m_{\psi,\rm{bare}}^2}{g|\dot{\sigma}_0|}\right) \,.
\end{eqnarray}

The energy density of $\psi$ particles when they decay into $\omega$ particles is roughly
given by $\rho_{\psi}=m_{\psi,\rm{eff}}n_{\psi}$ where the effective mass of the $\psi$
particles at the decay time is $m_{\psi,\rm{eff}}\simeq g\Phi$. We have neglected the
bare mass of $\psi$ because if the bare mass is significant compared to $g\Phi$ then
$n_{\psi}$ will be negligible due to the exponential suppression. For comparison the
energy density of the inflaton field if no preheating took place would be
$\rho_{\sigma}=m^2\sigma^2/2$. Therefore preheating is effective if
$\rho_{\psi}/\rho_{\sigma}\sim\mathcal{O}(1)$, and the ratio is given by
\bea\label{ip:ratio} \frac{\rho_{\psi}}{\rho_{\sigma}}&\simeq& \frac{1}{4\pi^3}
\left(\frac{\Phi}{m}\right)^{1/2}g^{5/2} \exp\left(-\pi
\frac{m_{\psi,\rm{bare}}^2}{g|\dot{\sigma}_0|}\right)   \nn \\  &=&
2.13g^{5/2}\left(\frac{10^{-6}\MPl}{m}\right)^{1/2}\exp\left(-\pi
\frac{m_{\psi,\rm{bare}}^2}{gm\Phi}\right)\,. \eea
We see that if $\psi$ has a negligible bare mass then preheating is effective for
$g=\mathcal{O}(1)$. However if the inflaton fields mass is suppressed $m\ll10^{-6}\MPl$ then
preheating can
also be efficient for $g\ll1$. It appears to be a coincidence that it is for the observationally
preferred value of $m\simeq 10^{-6}\MPl$ that $g=\mathcal{O}(1)$ leads to efficient preheating. This
constraint on $g$ is dependent on $m$ and the energy transfer is more
efficient for $m\ll 10^{-6}M_{\rm Pl}$, which would correspond to the inflaton
perturbation (\ref{ip:Pzetainf}) being smaller than the observed primordial perturbation.

\section{Nearly symmetric inflation}\label{ip:sec:ko}

In this section we consider the model proposed by Kolb et al.~\cite{Kolb:2004jm}. They
consider the case with two
inflaton fields, with similar mass. Assuming the minimum of the potential is at zero, a
generic form of the potential near the origin is \cite{Kolb:2004jm}
\begin{equation}\label{ip:V}
V(\phi_1,\phi_2)=\frac{m^2}{2}\left[\phi_1^2+\frac{\phi_2^2}{1+x}\right],
\end{equation}
where $0\leq x\ll1$ is the symmetry breaking parameter.

The basic idea of Ref.~\cite{Kolb:2004jm} is that the dominant cause of the density
perturbations may be due to an inhomogeneous reheating process caused by the slight
symmetry breaking. The background trajectory in the $\phi_1,\phi_2$ phase plane will be
nearly radial and passing close but not quite through the origin. There will also be
perturbations about this background trajectory. Those parallel to the path are adiabatic
perturbations \cite{Gordon:2000hv};
they correspond to time translations and will have no effect
on the point of closest approach. However the isocurvature perturbations, perpendicular
to the trajectory, will affect the closest approach.

The point of closest approach is of interest, since by an extension of (\ref{ip:npsi}),
assuming an interaction of the type (\ref{ip:interaction}) and $m_{\psi,\rm{bare}}=0$,
the comoving number density of $\psi$ particles is given by
\begin{equation}\label{ip:npsikofirst}
n_\psi=\frac{(g|\dot{\phi}_{\rm{min}}|)^{3/2}}{8\pi^3}\exp\left[-\frac{\pi g
|\phi_{\rm{min}}|^2}{|\dot{\phi}_{\rm{min}}|}\right],
\end{equation}
where $|\phi|=\sqrt{\phi_1^2+\phi_2^2}$. The subscript `$\rm{min}$' refers to quantities calculated at the time when the inflaton reaches its minimum value along its trajectory.


The value of $|\phi|$ and $|\dot{\phi}|$ at $t_{\rm{min}}$ are \cite{Kolb:2004jm},
\begin{eqnarray}\label{ip:phiko}
|\phi_{\rm{min}}|&=&\frac{\Phi\pi x }{\sqrt{8}}\sin(2\theta_0)\,, \\
\label{ip:phidotko} |\dot{\phi}_{\rm{min}}|&=&m\Phi\sqrt{1-x\sin^2{\theta_0}}\,.
\end{eqnarray}
As before the subscript $0$ refers to the initial values at the start of the preheating phase just
after the end of inflation, and $\tan\theta=\phi_1/\phi_2$.

The function $f(\theta_0)$, which we require to calculate $\zeta$, see (\ref{ip:zeta}),
is defined by \cite{Kolb:2004jm},
\begin{align}\label{ip:fthetafull}
\frac{\delta n_\psi}{n_\psi}&=f(\theta_0)\delta \theta_0 \nonumber
\\ &\approx -x\sin(2\theta_0)\left[
\frac{3}{4}+\frac{\pi^3\Phi gx}{8m} \left(4\cos(2\theta_0)+
\frac{1}{2}x\sin^2(2\theta_0) \right) \right]\delta\theta_0.
\end{align}
Assuming the primordial curvature perturbation is generated from the inflationary
isocurvature perturbations, the power spectrum of $\zeta$ from (\ref{ip:zeta}) and
(\ref{ip:fthetafull}) is
\begin{equation}\label{ip:Pzetafromtheta}
\mathcal{P}_\zeta=\alpha^2 f^2(\theta_0)\P_{\theta}=
\alpha^2f^2(\theta_0)\left(\frac{H_*}{2\pi\phi_*}\right)^2\,.
\end{equation}
Requiring that this has the required amplitude to match observations gives a constraint
on the four model parameters, $g,x,m,\theta_0$. We can greatly simplify $f(\theta_0)$ by
considering the parameter range that leads to efficient preheating.

The condition for efficient preheating (c.f.~(\ref{ip:ratio})), using
(\ref{ip:npsikofirst})--(\ref{ip:phidotko}) is
\begin{align}\label{ip:koefficiencypartial}
\frac{\rho_{\psi}}{\rho_{\sigma}}&=2.13g^{5/2}\left(1-x\sin(\theta_0)^2\right)^{3/4}
\left(\frac{10^{-6}\MPl}{m}\right)^{1/2}  \exp \left( -\frac{\pi^3\Phi
gx^2\sin(2\theta_0)^2}{8m\sqrt{1-x\sin(\theta_0)^2}}\right)  \nn  \\  &=\mathcal{O}(1)\,.
\end{align}
In order to satisfy this we require a large coupling, $g\gtrsim1$, and
$x^2\sin(2\theta_0)^2<m/\MPl\lesssim
10^{-6}$, so that there is no exponential suppression of the production of $\psi$
particles. If the symmetry is extremely weakly broken, $x\lesssim m/\MPl$ then from
Eqns.~(\ref{ip:zeta}) and (\ref{ip:fthetafull}) we have $\zeta\sim xm/\MPl$ which is much
smaller than observations require. Therefore we require $gx\MPl/m\gg 1$ and
$f(\theta_0)$, defined by (\ref{ip:fthetafull}), simplifies to
\bea\label{ip:fthetapartial} f(\theta_0)
=-\frac{\pi^3\Phi}{2m}gx^2\sin(2\theta_0)\cos(2\theta_0)\,. \eea
Substituting this into (\ref{ip:Pzetafromtheta}), taking $\alpha=1/4$ for definiteness
and requiring $\P_\zeta=2\times10^{-9}$ gives a constraint on the parameters from the COBE
normalisation,
\bea\label{ip:COBEko} gx^2\sin(2\theta_0)\cos(2\theta_0)=5.06\times10^{-4}\,. \eea
We substitute this constraint into the exponent of the efficiency of preheating, given by
(\ref{ip:koefficiencypartial}) to find that the exponent of the exponential is
$\mathcal{O}(-100\tan(2\theta_0)\MPl/(10^{6}m))$. We therefore require $|\sin(2\theta_0)|\ll1$ to
avoid an exponential suppression of $\psi$ particles. Substituting the constraint
equation (\ref{ip:COBEko}) into (\ref{ip:koefficiencypartial}) we find
\begin{align}\label{ip:preheatingko}
\frac{\rho_{\psi}}{\rho_{\sigma}}=2.13g^{5/2}\left(1-x\sin(\theta_0)^2\right)^{3/4}
\left(\frac{10^{-6}\MPl}{m}\right)^{1/2}  \exp \left(
-\frac{137\tan(2\theta_0)}{\sqrt{1-x\sin(\theta_0)^2}}\left(\frac{10^{-6}\MPl}{m}\right)
\right)\,. \end{align}

\subsection{Non-Gaussianity}\label{ip:sec:nongaussianity}

We can calculate the non-Gaussianity of the primordial curvature perturbation using
Eq.~(\ref{fNLformula}). If we define $\chi$ to be the angular field perturbations,
which generate the primordial curvature perturbation, then we have
$N_\chi=(1/|\phi_*|)N_\theta$ and $N_{\chi\chi}=(1/|\phi_*|^2)N_{\theta\theta}$. From
(\ref{ip:zeta}) we can identify $N=\log(n_{\psi})/4$. Therefore the non-linearity
parameter is given by
\begin{eqnarray}\label{ip:fNLdefnko}
f_{NL} =
\frac{5}{6}\frac{1}{\alpha}\left(\frac{n_\psi \frac{\partial^2
n_{\psi}}{\partial\theta^2}}{\left(\frac{\partial
n_{\psi}}{\partial\theta}\right)^2}-1\right).
\end{eqnarray}

From (\ref{ip:phiko}) and (\ref{ip:phidotko}) we see that the variation of
$|\dot{\phi}_{\rm{min}}|$ with respect to $\theta$ is suppressed by a factor of $m$
compared to $|\phi_{\rm{min}}|$. Hence the factor in front of the exponential in
(\ref{ip:npsi}) will give a negligible contribution to the non-Gaussianity and the
calculation simplifies considerably. Working with this approximation, from
(\ref{ip:npsikofirst}), (\ref{ip:phiko}) and (\ref{ip:fNLdefnko})
\bea\label{ip:fNLko} f_{NL}=-\frac{80}{3\pi^3}\frac{m}{\Phi}\frac{1}{gx^2}
\frac{\cos(4\theta_0)}{\sin(4\theta_0)}\,. \eea

Substituting in the constraint (\ref{ip:COBEko}) and working in the regime that
$|\sin(2\theta_0)|\ll1$, which is required for efficient preheating,  $f_{NL}$ simplifies to
\bea\label{ip:fNLkosimple} f_{NL}\simeq -0.012
\frac{1}{\sin(2\theta_0)}\left(\frac{m}{10^{-6}\MPl}\right)\,. \eea
Note that this is the inverse of the exponent of the exponential in
(\ref{ip:preheatingko}) up to a factor of order unity. $|f_{NL}|$ is less than
one except for extremely small $|\sin(2\theta_0)|$. Also note that $\fNL$ is negative in this model.

%
%

Our result for $f_{NL}$ is not the same as the result in \cite{Kolbx2}. They perform a
calculation at leading order in $x$ and find that $\fNL$ can be large for most values of
$\theta_0$. However this gives the correct result only when
$x\MPl/m\lesssim1$, which requires $x\lesssim 10^{-6}$. For the parameter values which
lead to efficient reheating and angular perturbations which give the dominant
contribution to the primordial curvature perturbation this approximation cannot be used.

To be explicit, from (\ref{ip:npsikofirst}) we find (Eq.~(11) in \cite{Kolbx2}),
\be\label{ip:eq11ko} \frac{\delta
n_{\psi}}{n_{\psi}}=\frac32\frac{\delta|\dot{\phi}_{\rm{min}}|}{|\dot{\phi}_{\rm{min}}|}
+\frac{\pi g|\phi_{\rm{min}}|^2}{|\dot{\phi}_{\rm{min}}|}
\frac{\delta|\dot{\phi}_{\rm{min}}|}{|\dot{\phi}_{\rm{min}}|} -\frac{2\pi
g|\phi_{\rm{min}}|^2}{|\dot{\phi}_{\rm{min}}|}
\frac{\delta|\phi_{\rm{min}}|}{|\phi_{\rm{min}}|}\,, \ee
and \hspace{\stretch{1}} the\hspace{\stretch{1}} three\hspace{\stretch{1}} terms
\hspace{\stretch{1}}are\hspace{\stretch{1}} respectively\hspace{\stretch{1}}
$\mathcal{O}(x\sin(2\theta_0))$,\hspace{\stretch{1}}
$\mathcal{O}(gx^2\sin^3(2\theta_0)\MPl/m)$\hspace{\stretch{1}} and \\
 $\mathcal{O}(gx^2\sin(2\theta_0)\cos(2\theta_0)\MPl/m)$.
We see that the first term is leading order in an expansion in $x$ and \cite{Kolbx2}
calculates the non-Gaussianity arising from this term, but for the parameter range we are
interested in the third term, which is the only term when one neglects the variation of
$|\dot{\phi}_{\rm{min}}|$ with respect to $\theta$, dominates over both other terms by
around five orders of magnitude. This is because we have $m<10^{-6}\MPl$ and the last term is
multiplied  by a factor of $\MPl/m\gg x^{-1}>1$ compared to the other two terms. It is the
third term of (\ref{ip:eq11ko}) that we have used to calculate (\ref{ip:fNLko}).

\subsection{Parameter constraints}

Because the model has four parameters, three of which are free after taking account of
the COBE constraint (\ref{ip:COBEko}) it is not easy to plot the allowed parameter range.
We will plot the allowed range of $g$ and $\theta$ for a given choice of $m$. The bounds
also depend on the theoretical cut-off we put on the parameters. Since we have made an
expansion in $x$ and taken the leading order terms in $|\phi_{\rm{min}}|$ and
$|\dot{\phi}_{\rm{min}}|$, see (\ref{ip:phiko}) and (\ref{ip:phidotko}), we will require
$x\lesssim 0.1$ so that the results are accurate to about $10\%$. For definiteness we
require $\rho_{\psi}/\rho_{\sigma}\geq0.1$ as the condition for efficient preheating.

\begin{figure}
\scalebox{0.7}{\includegraphics*{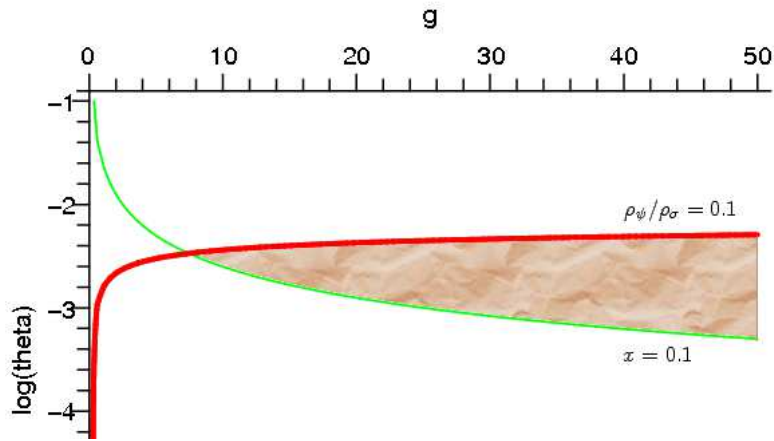}} \caption{The allowed parameter range of $g$ and
$\log_{10}(\theta)$ for $m=10^{-7}\MPl$. The rising thick red line is
$\rho_{\psi}/\rho_{\sigma}=0.1$ with larger values below this line and the falling thin green line
is $g\theta=0.00025$ which is the COBE constraint for $x=0.1$. Values with $x<0.1$ lie above this
line. The shaded, textured area is the allowed region of parameter space.}\label{fig:m7crop}
\end{figure}

The minimum possible value of $g$ increases as $m$ is reduced. For $m/\MPl=10^{-6},
10^{-7}$ and $10^{-8}$ we have from (\ref{ip:COBEko}) and (\ref{ip:preheatingko})
$g_{\rm{min}}=1.6,7.6$ and $47$ respectively. A value of $g\gg1$ is unattractive for
theoretical reasons, since loop corrections to the effective potential are then likely to
large. We require $m\lesssim10^{-6}\MPl$ so that the inflationary adiabatic perturbations
are suppressed. The allowed range of $g$ and  $\theta$ is shown in Fig.~\ref{fig:m7crop} for
$m=10^{-7}\MPl$. 

Fig.~\ref{fig:m7crop} is for $\theta\simeq0$, there is a second branch of allowed values at
$\theta\simeq \pi/2$, since in both of these regions we have $|\sin(2\theta_0)|\ll1$. The
allowed region is very similar for the second branch near $\theta_0=\pi/2$ but slightly
reduced because the term $(1-x\sin(\theta_0))$ is slightly smaller in the branch
$\theta_0=\pi/2$ and this makes preheating slightly less efficient, see
(\ref{ip:preheatingko}).

The allowed range of $\theta_0$ is small, approximately we require
$0.001\lesssim\theta_0\lesssim0.01$ even for large $g$. Therefore we require a significant
fine tuning of the initial conditions so that $\phi_2$ is very small initially. 

Because there is a minimum allowed $\theta_0\gtrsim 0.001$ to satisfy the COBE constraint
(\ref{ip:COBEko}) we see from (\ref{ip:fNLkosimple}) that $|\fNL|\lesssim1$ in the allowed
parameter range. Hence the perturbations from this model are quite close to Gaussian.

Finally we give the spectral index for this model, as calculated in Sec. IV D of \cite{byrnes2} upto
second order in slow roll,
\bea\label{nzetako} n_\zeta-1=-2x\epsilon \cos (2\theta_0)-\frac{10}{3}\epsilon^2. \eea
Hence the perturbations are slightly red but very close to scale invariant. Since $x<1$ it is
possible for the second-order in slow-roll term to dominate over the leading order result.

\section{Non symmetric inflation}\label{ip:sec:ma}

In this section we consider a model where the symmetry is strongly broken during
inflation, $m_1\gg m_2$ in (\ref{ip:V2field}), and the isocurvature perturbations during
inflation generate the primordial density perturbation. This model was suggested by
Matsuda \cite{Matsuda:2006ee}. However the motivation in this paper was to look for
models with a low energy scale of inflation and the models considered in detail either
had an additional period of inflation or effects from brane world models of inflation.
Here we consider the most economical model, with the potential (\ref{ip:V2field}) valid from the
time of Hubble exit of the observable scales till instant preheating. We calculate the constraints
on the parameter of this model and calculate the non-Gaussianity.

Assuming that $\phi_2\lesssim\phi_1$, we have $V\simeq m_1^2\phi_1^2/2$ during inflation
because $m_1\gg m_2$. From the equations of motion in the
slow roll limit, $3H\dot{\phi}_I+m^2_I\phi_I\simeq0$ we can estimate
\bea \frac{\dot{\phi}_2}{\dot{\phi}_1}=\left(\frac{m_{2}}{m_1}\right)^2\,. \eea
If $m_1=10m_2$ then $\phi_2$ will only roll $1\%$ of the distance that $\phi_1$ rolls
during slow-roll inflation. Therefore it is a good approximation to treat the background
trajectory as straight during inflation with $\phi_2$ a constant. Because the background
trajectory is along $\phi_1$ we can identify the adiabatic perturbation as
$\delta\sigma=\delta\phi_1$ and the isocurvature perturbations $\delta\chi=\delta\phi_2$.
Then the total field velocity is given by $\dot{\phi}_1$ and the minimum distance from
the origin of the field trajectory is $\phi_2=\phi_{2*}$, which occurs at the time when
$\phi_1=0$ for the first time after inflation ends. Assuming that $m^2_{\psi,\rm{bare}}=
g^2\phi_2^2<gm\Phi$, we have from (\ref{ip:npsi}) \cite{Matsuda:2006ee},
\bea\label{ip:npsipartialma} n_{\psi}= \frac{\left(gm_1\Phi\right)^{3/2}}{8\pi^3}
\exp\left(-\pi\frac{g\phi_2^2}{m_1\Phi}\right)\,, \eea
where as before we have used $|\dot{\phi}_1|=m_1\Phi$, which follows from conservation of
energy if we neglect the expansion of the universe during the first half oscillation of
the inflaton field.

The slow-roll parameters for this model are given by
\bea \e=\esigsig=\frac{\MPl^2}{4\pi}\frac{1}{\phi_1^2}, \qquad \eta_{\sigma\chi}=0, \qquad
\eta_{\chi\chi}=\frac{\MPl^2}{4\pi}\frac{m_2^2}{m_1^2}\frac{1}{\phi_1^2}\ll\e\,, \eea
and the higher order slow roll parameters are zero in the limit of a straight background
trajectory. Throughout this paper we use the notation that $\sigma$ is the adiabatic field and $\chi$ is the isocurvature field \cite{Gordon:2000hv}.

Since $\e\simeq0.008$ for this model, if the inflationary isocurvature
perturbation is converted into the primordial curvature perturbation then at
leading order in slow roll we have \cite{WBMR}
\bea\label{ip:tiltma} n_{\zeta}-1\simeq-2\e\simeq-0.016. \eea
Therefore the power spectrum is quite close to scale invariant but not as close as the
previous model where the potential was nearly symmetric, see (\ref{nzetako}).

\begin{figure}
\scalebox{0.7}{\includegraphics*{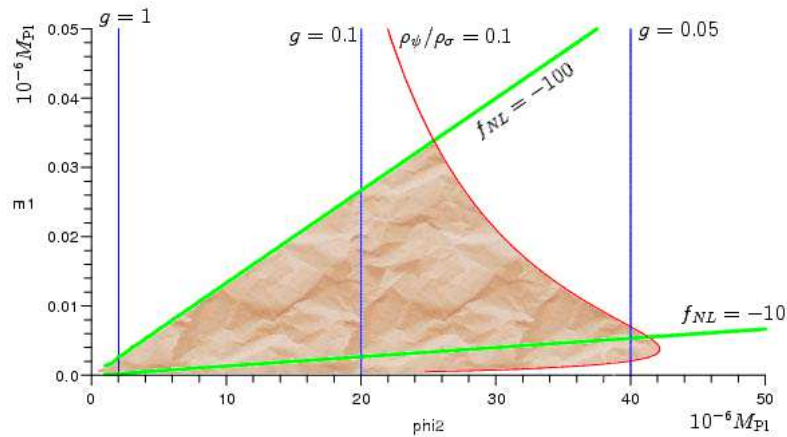}}\caption{The allowed parameter range of
$\phi_2$ and $m_1$. The curved red line is
$\rho_{\psi}/\rho_{\sigma}=0.1$ with larger values to the left of the curve. The straight thick
green lines mark constant values of $\fNL$ as marked on the diagram. The shaded, textured area
corresponds to
the allowed region where all of the constraints are satisfied. The vertical, blue lines mark values
of
constant $g$ as marked in the diagram.}\label{fig:matsuda}
\end{figure}

From (\ref{ip:zeta}) and (\ref{ip:npsipartialma}) the primordial curvature perturbation
is given by,
\bea \zeta=-2\pi\alpha\frac{g\phi_2}{m_1\Phi}\delta\phi_2\,. \eea
Using $\P_{\phi_2}=Nm_1^2/(6\pi^2)$, which follows from
$\epsilon=\MPl^2/(4\pi\phi_1^2)\simeq1/(2N)$,
and $V=3\MPl^2H^2$ in the slow-roll limit, taking $N=60$
and $\alpha=1/4$ for definiteness (the constraints hardly change if we instead take $\alpha=1/3$),
the COBE normalisation $\P_\zeta=2\times10^{-9}$ gives
\bea\label{ip:COBEma} g\phi_2=2.0\times10^{-6}\MPl\,. \eea
Substituting this into (\ref{ip:npsipartialma}) we find
\bea\label{ip:npsima}  n_{\psi}= \frac{\left(gm_1\Phi\right)^{3/2}}{8\pi^3}
\exp\left(-90\phi_2\left(\frac{10^{-6}}{m_1}\right)\right)\,. \eea
We can calculate the non-Gaussianity in a similar way to the previous section
\ref{ip:sec:nongaussianity}, from (\ref{ip:fNLdefnko}) and (\ref{ip:npsipartialma}) this gives
\bea f_{NL}=\frac{5}{3\pi}\frac{1}{g}\frac{m_1\Phi}{\phi_2^2}\,, \eea
and substituting in the constraint (\ref{ip:COBEma}) this simplifies to
\bea\label{ip:fNLma} f_{NL}=-0.075\frac{1}{\phi_2}\left(\frac{m_1}{10^{-6}}\right)\,.
\eea
Note that again $\fNL$ is negative in this model, and that up to a numerical factor of order unity
this is just the inverse of the exponent in the
exponential of Eq.~(\ref{ip:npsima}). The WMAP bound that $|f_{NL}|<100$
\cite{wmap5}\footnote{In fact the ``headline'' constraint on $\fNL$ prefers a positive
value but the constraint from Minkowski functionals using the same data favours a negative $\fNL$.
Here we use an approximate and reasonably conservative bound. In Fig.~\ref{fig:matsuda} one can see
how the constraint on the parameters tighten if one uses a lower bound of $\fNL>-10$ as opposed
to $\fNL>-100$.} requires
\bea\label{ip:phi2minma} \phi_2>7.5\times10^{-4}\left(\frac{m_1}{10^{-6}}\right)\,. \eea
Note that since $\delta\phi_{2*}=\mathcal{O}(H_*)=\mathcal{O}(m_1)$, the condition for nearly
Gaussian
perturbations is equivalent to the requirement $\delta\phi_2\ll\phi_2$.

From (\ref{ip:ratio}) the efficiency of preheating is given by
\bea
\frac{\rho_{\psi}}{\rho_{\sigma}}=2.13g^{5/2}\left(\frac{10^{-6}\MPl}{m_1}\right)^{1/2}
\exp\left(-90\phi_2\left(\frac{10^{-6}}{m_1}\right)\right)\,. \eea

The parameter $m_2$ turns out to be irrelevant, provided it satisfies $m_2\ll m_1$.
Therefore we really have a three parameter model, $m_1,g$ and $\phi_2$ and one constraint
relating $g$ and $\phi_2$ given by (\ref{ip:COBEma}). Therefore there are only two free
parameters, which we plot in Fig.~\ref{fig:matsuda}. We see that only a small parameter
range is allowed. Roughly, we require $m_1\lesssim3\times10^{-8}\MPl$ and
$\phi_2<4\times10^{-5}\MPl$. This is consistent with our initial assumption that
$m_1^2\phi_1^2\gg m_2^2\phi_2^2$. For much of the allowed parameter range the
non-Gaussianity is significant, $f_{NL}<-1$. 


\section{Symmetric inflation followed by asymmetric preheating}\label{ip:sec:by}

In this section we consider a model with a symmetric potential, $m=m_1=m_2$ in (\ref{ip:V2field}),
but unlike in the previous two models we break the symmetry during preheating, $g_1\neq g_2$
in (\ref{ip:g2field}). This section is based on Byrnes and Wands \cite{byrnes}. Because the
parameter constraints for this model have already been calculated in \cite{byrnes} we here just
summarise the results from this model. We will see that this model requires a remarkably similar
degree of fine tuning to the initial trajectory as the previous two models, even though here the
method of breaking the symmetry between the two fields is very different from the previous two
models considered.

This model has four parameters, the inflaton mass $m$, the two coupling constants $g_1$
and $g_2$ defined by (\ref{ip:g2field}) and the angle of the background trajectory
$\theta$ defined by $\tan\theta=\phi_1/\phi_2$. Unlike in the previous two models
discussed $\theta$ is a constant, because the potential is symmetric.

The formula for the number of $\psi$ particles produced during instant preheating, shown
in (\ref{ip:npsi}), still holds if we replace the single field coupling constant $g$ with
the effective coupling constant \cite{byrnes}
\begin{equation}\label{ip:gtilde}
\tilde{g}^2(\theta) \equiv g_1^2\cos^2\theta+g_2^2\sin^2\theta\,.
\end{equation}
Because the straight background trajectory of this model goes through the origin where
$V=0$ we can choose $m_{\psi,\rm{bare}}=0$. Then we have, see (\ref{ip:npsi})
\begin{eqnarray}\label{ip:npsiko}
n_\psi= \frac{1}{8\pi^3} \left(\tilde{g}|\dot{\sigma}_0|\right)^{3/2}\,.
\end{eqnarray}
The ratio of the energy density of the $\psi$-field to that of the inflaton field at the
point when the inflaton would be at its maximum after the first explosive creation of
$\psi$ particles is $\rho_\psi/\rho_\sigma\sim 2\tilde{g}^{5/2}(10^{-6}M_{\rm
Pl}/m)^{1/2}$, see (\ref{ip:ratio}). Thus if $\tilde{g}>1$ we see that most of the
inflaton's energy density is transferred to the preheat field during the first
oscillation for $m\sim 10^{-6}M_{\rm Pl}$. This constraint on $\tilde{g}$ is dependent on
$m$ and the energy transfer is more efficient for $m\ll 10^{-6}M_{\rm Pl}$, which would
correspond to the inflaton perturbation (\ref{ip:Pzetainf}) being smaller than the
observed primordial perturbation.

Assuming the decay products are massive, non-relativistic particles, then the linear
metric perturbation (\ref{ip:zeta}) is given by $\zeta=(1/3)\delta n_\psi/n_\psi$. Thus
if the effective coupling, $\tilde{g}$, is dependent upon the phase of the complex field
we have
\begin{equation}\label{ip:preheatzeta}
\zeta =
\frac{F(R,\tan\theta)}{2} \frac{\delta\chi}{\sigma} \,,
\end{equation}
where from (\ref{ip:gtilde}) and (\ref{ip:npsiko}) it follows that
\begin{equation}\label{ip:Fby}
F(R,\tan\theta) \equiv \frac{\tan\theta(1-R^2)}{R^2+\tan^2\theta}
 \quad \mathrm{and} \quad
R \equiv \frac{g_1}{g_2}\,.
\end{equation}
We only need to consider the case $R<1$ and $0<\theta<\pi/2$, due to the symmetries of
this model. The power spectrum is then given from $\P_\chi=\sigma^2/(3/\pi)(m/\MPl)^2$ and
(\ref{ip:preheatzeta}) by
\begin{equation}
\label{ip:Pzetaiso} \mathcal{P}_{\zeta,\textrm{iso}}
\simeq \frac{F^2(R,\tan\theta)}{12\pi} \left( \frac{m}{M_{Pl}} \right)^2 \,.
\end{equation}

Comparing this with Eq.~(\ref{ip:Pzetainf}) we see that either
$\mathcal{P}_{\zeta,\textrm{iso}}$ or $\mathcal{P}_{\zeta,\textrm{inf}}$ can dominate the
primordial density perturbation depending on the values of $R$ and $\theta$. The
asymmetric preheating creates the dominant primordial perturbation if
\begin{equation}
F(R,\tan\theta) > 4N \simeq 240 \,.
\end{equation}
From (\ref{ip:Fby}) we require $R<1/480$ for the isocurvature perturbations to dominate
in any range of $\theta$. If $R\ll1/480$ then the isocurvature perturbations dominate for
$240R^2<\theta<1/240$, while larger $R$ will have a smaller range of suitable $\theta$.
The range of $\theta$ where isocurvature perturbations dominate will be larger if $N<60$.
We require that the inflaton trajectory is almost exactly along $\phi_1$, the field with
the smaller coupling constant $g_1$. 

The COBE normalisation that $\mathcal{P}_\zeta\simeq2\times10^{-9}$ constrains the three
free parameters $m$, $R$ and $\theta$. Perturbations produced at preheating from
isocurvature field fluctuations are of the required size if
\begin{equation}
F(R,\tan\theta)=200 \left( \frac{10^{-6}M_{Pl}}{m} \right) \,.
\end{equation}

We can calculate the non-Gaussianity in a similar way to the previous two models, the
result is\footnote{Note that we have used the opposite sign
convention for $f_{NL}$ used in \cite{byrnes} in order to be consistent with the WMAP convention.}
\begin{eqnarray}
f_{NL}=\frac56\frac{N_{,\theta\theta}}{N_{,\theta}^2}
= \frac{5}{3(1-R^2)} \frac{(R^2-\tan^2\theta)}{\sin^2\theta} \,.
\end{eqnarray}
This is of order unity, and hence likely to be unmeasurable, for $\theta\sim R$, but can
become large for smaller $\theta$. For example if $R=10^{-3}$, then the perturbations
from asymmetric preheating dominate in the range $0.00026<\theta<0.0039$ and $f_{NL}$ has
its maximum absolute value of $24$ at $\theta=0.00026$, but is far smaller for most of
the allowed range of $\theta$. Unlike in the two previous models this model predicts a positive
value of $\fNL$.

Finally we comment on the spectral index. At first order in slow roll the spectral index is scale
invariant, and to second order in slow roll \cite{byrnes}
\begin{eqnarray}\label{nzetasymm}
n_\zeta-1 = -\frac{10}{3}\epsilon^2\simeq-0.0002\,.
\end{eqnarray}
This is extremely close to scale invariant and unlikely to be observationally distinguishable.
The result is also a special case of the first model we considered with a nearly symmetric
potential in Sec.~\ref{ip:sec:ko} with $x=0$.

\section{Conclusion}\label{ip:sec:conclusion}

We have considered three models for converting the inflationary isocurvature perturbation
into the primordial curvature perturbation. For two models where the symmetry is broken during
inflation \cite{Kolb:2004jm,Matsuda:2006ee} we have calculated the allowed parameter space, subject
to the constraints that instant 
preheating is efficient, that the inflationary isocurvature perturbation generates the primordial
curvature perturbation and that the amplitude of perturbations satisfies the COBE normalisation.
For a model where the symmetry is broken during instant preheating we have summarised the
constraints found in a previous paper for completeness \cite{byrnes}.

It turns out that all of the models require
some fine tuning to be effective. We can understand this by considering how the total
power spectrum is generated. From (\ref{Pzetasplit})
the first term, coming from the inflationary adiabatic perturbations are always given by
(\ref{ip:Pzetainf}) and $\partial
N/\partial\sigma=\mathcal{O}(\e^{-1/2})$. Hence the primordial power spectrum is boosted by a
factor $\epsilon^{-1}$ relative to the inflaton field perturbations. If primordial
density perturbations due to isocurvature field perturbations are to dominate, then their
effect on the expansion history must be boosted by a larger factor. In contrast with the
inflaton, the amplitude of the density perturbations resulting from orthogonal
perturbations is dependent upon the physics after inflation. In order for the
inflationary isocurvature perturbations to generate the dominant part of the primordial
curvature perturbation we require
\begin{equation}\label{ip:Nchi}
\left( \frac{\partial N}{\partial \chi} \right)^2 \gg \left( \frac{\partial N}{\partial
\sigma} \right)^2=\mathcal{O}\left(\frac{1}{\e}\right)=\mathcal{O}(100)\,.
\end{equation}
It turns out to be hard to satisfy this condition except by fine tuning the parameters.
We also need to choose parameters that lead to efficient preheating, (\ref{ip:ratio}),
which often means that the coupling must be strong, $g\gtrsim1$ and that the mass of the
$\psi$ particles must be small at the time of their creation. Finally there is also the
observational constraint that the non-Gaussianity must not be too large.

The parameter constraints are approximately that for the nearly symmetric potential we
required $10^{-8}\lesssim m/\MPl<10^{-6}$ and strong coupling, $g\gtrsim1$; for the
strongly broken potential we require the mass hierarchy $m_2\ll
m_1\lesssim3\times10^{-8}$ and an even stronger coupling, $g\gtrsim10$; and for the non-symmetric
reheating model we require a large
ratio of the coupling constants $g_1/g_2\gtrsim500$. Perhaps more serious than the
constraints on the parameters are that for each model we require fine tuning of the
initial conditions so that the background trajectory lies very close to one of the axes,
i.e.~we require that the initial value of one of the two inflaton fields $\phi_I$
is very small. Unless the broken symmetry of the models can play a role in creating these
initial conditions the fine tuning is unattractive.

The size of the isocurvature perturbations in each model is given by
$\delta\chi=\mathcal{O}(H)=\mathcal{O}(m_1)$, independent of the background trajectory. It is only
when the
background trajectory lies very close to $\chi=0$ that these small perturbations
correspond to a significant perturbation in the value of $\chi$, i.e.~it is only for very
small $\chi$ that $\delta\chi/\chi$ is non-negligible. This is what we require in order
to satisfy the inequality (\ref{ip:Nchi}) for all of the models we have considered.
However we can not have a too small value of $\chi$ because if $\chi=\mathcal{O}(\delta\chi)$ then
the non-Gaussianity is very large and this is ruled out by observations, see the comment
after (\ref{ip:phi2minma}).

It is interesting to note that the fine tuning of the initial trajectory to be close to one of the
inflaton fields axes is also required in several other models in order for them to generate a large
non-Gaussianity. Examples include an inhomogeneous end of inflation
\cite{alabidi:endinf,sasaki:endinf,Salem:2005nd} and during multiple-field slow-roll inflation
\cite{byrnes5}.
\newline

Ultimately, observational data will determine whether the primordial density perturbation
has a significant deviation from scale-invariance, $n_{\zeta}-1\simeq-0.03$, as predicted by
adiabatic fluctuations in the inflaton field driving chaotic inflation. An equally
important observable is the amplitude of the primordial gravitational wave background
that is predicted. The amplitude of gravitational waves is determined directly by the
energy scale during inflation and is given by \cite{BTW}
\begin{equation}
{\cal P}_T = \frac{32}{3\pi}\left( \frac{m}{M_{\rm Pl}} \right)^2 N\,.
\end{equation}

Adopting the COBE normalisation for ${\cal P}_\zeta\simeq 2\times10^{-9}$,
the primordial tensor-scalar ratio is given by
\begin{equation}
r \equiv \frac{{\cal P}_T}{{\cal P}_\zeta} \simeq 10^{-1}\left(\frac{m}{10^{-6}
M_{Pl}}\right)^2\,.
\end{equation}
Hence we see that the tensor-scalar ratio is reduced if $m<10^{-6}M_{Pl}$. Thus if
isocurvature field fluctuations contribute significantly to the primordial density
perturbation the gravitational wave background is much smaller than if the density
perturbations are due solely to inflaton perturbations \cite{Sasaki:1995aw,WBMR,BTW}. If
the isocurvature perturbations have no effect towards generating the primordial curvature
perturbation then $r=16\e\simeq0.13$. If $r$ is this large then the tensor background should be
detected with the Planck experiment \cite{planckbluebook}.

If the primordial curvature perturbation is generated from the inflationary isocurvature
perturbation then the spectral index depends on the model of inflation, but in each case
it is closer to scale invariance than the curvature perturbations during inflation. The
model with a non-symmetric potential has the largest deviation from scale invariance, as
given by (\ref{ip:tiltma}), while the nearly symmetric potential is very close to scale
invariance (\ref{nzetako}), which for the allowed
parameter range satisfies $-0.001<n_{\zeta}-1<0$. The model with a symmetric potential is
closest of all to scale invariance, shown by (\ref{nzetasymm}), with the tilt being
second order in slow-roll parameters $n_{\zeta}-1\simeq-0.0002$. Observationally
this model is therefore likely to be indistinguishable from a Harrison-Zel'dovich
spectrum, $r=n_\zeta-1=0$. This is mildly disfavoured by observations \cite{wmap5}.

For each model we have also calculated the non-Gaussianity, specifically the bispectrum
parameterised by $f_{NL}$. Only in the model with a non-symmetric potential is
$|f_{NL}|>1$ for a significant range of the viable parameter range, see
Fig.~\ref{fig:matsuda}. In this model $\fNL<0$ so if the hint of a detection of $\fNL>0$
\cite{Yadav:2007yy} is confirmed then this model will be ruled out. For the model with a nearly
symmetric potential $|f_{NL}|\lesssim1$ for
the entire parameter range, while for the model with non symmetric preheating it is
possible but not preferred to have $f_{NL}>1$. 

If instant preheating is completely efficient then the inflaton field will completely
decay into $\psi$ particles around the first time $\sigma$ reaches its minimum after
inflation ends and all isocurvature perturbations will be washed out. However if the
process is less efficient then it is possible that a residual fraction of isocurvature
perturbations will survive until today. The amount will depend on the subsequent
reheating processes, but in the limit that the primordial curvature perturbation is
entirely generated from the inflationary isocurvature perturbations then the primordial
curvature and isocurvature perturbations will be totally correlated. Constraints on
the fraction of isocurvature perturbations in the CMB are given in
\cite{wmap5,Bean:2006qz,Trotta:2006ww,Keskitalo:2006qv}. However if there is
an isocurvature perturbation present after preheating then it may affect the subsequent evolution
of the primordial curvature perturbation, which is otherwise conserved on large scales
\cite{WMLL,Choi:2008et}.

\acknowledgements

CB thanks Robert Crittenden, Andrew Liddle and David Wands for useful discussions on this topic
and the Institute of Cosmology and Gravitation, University of Portsmouth for hospitality during
several visits where part of this
work was completed. CB acknowledges financial support from the Deutsche Forschungsgemeinschaft.



\end{document}